\documentstyle[preprint,aps]{revtex}
\begin{document}
\draft
\preprint{CALT-68-1959}
\title{Semileptonic Decays of Heavy Tetraquarks}
\author{Chi-Keung Chow}
\address{California Institute of Technology, Pasadena, CA 91125}
\date{\today}
\maketitle
\begin{abstract}
There exist heavy tetraquark states $QQ \bar q \bar q$ in the heavy quark
limit.
These states are stable with respect to strong interactions and hence must
decay weakly.
It is shown that the semileptonic decay depends on a single Isgur--Wise form
factor, which can be expressed in term of the Isgur--Wise form factors which
govern the semileptonic decays of the $Qqq$ and $QQq$ baryons.
\end{abstract}
\narrowtext
The possibility of the existence of exotics, i.e, hadrons with properties
incompatible with a $q\bar q$ or $qqq$ description, is an important issue in
QCD.
In particular, the existence of tetraquark (or di-meson) states has been
proposed by Jaffe \cite{T1} and since then investigated by many other groups
\cite{T2,T3,T5,T6}.

The case of binding two heavy quarks to two light antiquarks to form a heavy
tetraquark $QQ \bar q \bar q$ is especially interesting.
The two heavy quarks are bound by the short range Coulumbic chromoelectric
attraction in the $3 \otimes 3 \rightarrow \bar 3$ channel.
The bound state will have binding energy of order $E\sim\alpha_s^2(m_Q)m_Q$.
When $m_Q$ is so large such that $E\gg\Lambda_{\rm QCD}$, the large binding
energy forbids the dissociation $QQ \bar q \bar q\rightarrow Q \bar q
+ Q \bar q$.
Moreover, the light degrees of freedom cannot resolve the closely bound $QQ$
system, which has size of order $(m_Q\,\alpha_s(m_Q))^{-1}\ll\Lambda_{\rm
QCD}^{-1}$.
This results in bound states that have ``brown mucks'' similar to those of
$\bar\Lambda_Q$ states, with $QQ$ playing the role of the heavy antiquark.
Hence the stability of $\bar\Lambda_Q$ implies that $QQ \bar q \bar q$ is also
safe from decaying through $QQ \bar q \bar q \rightarrow QQq
+ \bar q \bar q \bar q$.
As the result, $QQ \bar q \bar q$ is stable with respect to strong
interactions and must decay weakly.

Since the ``brown mucks'' of the ground state tetraquarks are spinless, the
spins of the tetraquarks are just given by the spins of the heavy quarks.
As the result, the ground states are degenerate spin $S=0$ or 1 tetraquarks.
We will denote the tetraquarks by $|T(p,S,m)\rangle$ where $m$ is the
$z$-component of the spin.
The normalizations of these states are given by
\begin{equation}
\langle T(p',S',m')|T(p,S,m)\rangle = 16\pi^3 \delta_{SS'} \delta_{mm'}
\delta^3({\bf p}-{\bf p'}),
\end{equation}
such that, when the masses of the tetraquarks go to infinity, the states are
still well defined.

Since the two heavy quarks are bound by a Coulumbic potential, we can
represent the heavy degrees of freedom as
\begin{equation}
|Q_a Q_b (v,S,m)\rangle = \int d^3{\bf p} \, |Q_a (v_a,s_a)\rangle \otimes |Q_b
(v_b,s_b) \rangle \otimes \psi(B;{\bf p}) \, (\textstyle {1\over2},s_a;
\textstyle {1\over2},s_b|S,m).
\end{equation}
The Clebsch--Gordon coefficient describes the spin structure of the tetraquark,
and ${\bf p}=m_a\,({\bf v}-{\bf v_a})=-m_b\,({\bf v}-{\bf v_b})$ is the
relative momentum between the heavy quarks.
$\psi(B;{\bf p})$ is the ground state Coulumbic wavefunction in momemtum space
\begin{equation}
\psi(B;{\bf p})={4B^{5/2}\over ({\bf p}^2+B^2)^2} ,
\end{equation}
with $B=\mu_{ab}\,\alpha_s(\mu_{ab})$ the reciprocal of the Bohr radius.
Together with the spinless light degrees of freedom $|\phi(v) \rangle$, we get
the decomposition
\begin{eqnarray}
|T_{ab}(v,S,m)\rangle &=& |Q_a Q_b \bar q \bar q (v,S,m)\rangle \nonumber\\
&=& \int d^3{\bf p} \, |Q_a (v_a,s_a)\rangle \otimes |Q_b (v_b,s_b) \rangle
\otimes \psi(B;{\bf p}) \otimes |\phi(v) \rangle \, (\textstyle {1\over2},s_a;
\textstyle {1\over2},s_b|S,m).
\label{dec}
\end{eqnarray}

We define the Isgur--Wise form factor $\tilde \eta_{abc} (w)$ of the
semileptonic $T_{ab} \rightarrow T_{ac}$ decay by
\begin{eqnarray}
\langle T_{ac}(v',S',m')&|&\bar Q_c \Gamma Q_b|T_{ab}(v,S,m)\rangle \nonumber\\
& &= \tilde \eta_{abc}(w) \, \delta_{s_as_a'} \, \bar u_c(v', s'_c) \Gamma
u_b(v, s_b) \, (\textstyle {1\over2},s_a; \textstyle {1\over2},s_b|S,m) \,
(\textstyle {1\over2},s_a'; \textstyle {1\over2},s_c'|S',m'),
\label{def}
\end{eqnarray}
where $w=v\cdot v'$.
The form factor $\tilde\eta_{abc}(w)$ contains contributions from both
perturbative QCD, which describes the attraction between the two heavy quarks,
and non-perturbative QCD, which accounts for the interaction of the $QQ$ system
with the light degrees of freedom.

By Eq. (\ref{dec}), the matrix element can also be decomposed as
\begin{eqnarray}
\langle T_{ac}(v',S',m')|&\bar Q_c& \Gamma Q_b|T_{ab}(v,S,m)\rangle \nonumber\\
&=& \int d^3{\bf p'} \int d^3{\bf p} \,\psi(C;{\bf p'}) \psi(B;{\bf p}) \,
\langle \phi (v')|\phi (v) \rangle \, \langle Q_a (v_a',s_a')|Q_a (v_a,s_a)
\rangle \nonumber\\ & &\quad \langle Q_c (v_c',s_c') |\bar Q_c \Gamma Q_b|Q_b
(v_b,s_b) \rangle \, (\textstyle {1\over2},s_a; \textstyle {1\over2},s_b|S,m)
\, (\textstyle {1\over2},s_a'; \textstyle {1\over2},s_c'|S',m'),
\end{eqnarray}
where analogously ${\bf p'}=m_a\,({\bf v'}-{\bf v'_a})=-m_c\,({\bf v'}-{\bf
v'_c})$ and $C=\mu_{ac}\,\alpha_s(\mu_{ac})$.

In order the evaluate $\langle \phi (v')|\phi (v) \rangle$, recall that
\begin{equation}
|\bar\Lambda_Q (v,s)\rangle = |\bar Q (v,s)\rangle \otimes |\phi (v)\rangle .
\end{equation}
Then
\begin{eqnarray}
\langle\bar\Lambda_c(v',s')|\bar Q_b \Gamma Q_c|\bar\Lambda_b(v,s)\rangle &=&
\langle \phi (v') |\phi (v) \rangle \,\langle \bar Q_c (v',s') |\bar Q_b \Gamma
Q_c|\bar Q_b (v,s) \rangle \nonumber\\ &=& \langle \phi (v')|\phi (v) \rangle
\, \bar v_c (v',s') \Gamma v_b (v,s).
\end{eqnarray}
The Isgur--Wise form factor $\eta(w)$ of $Qqq$ baryons are defined by
\cite{B4,B5,B6,B7}
\begin{equation}
\langle\bar\Lambda_c(v',s')|\bar Q_b \Gamma Q_c|\bar\Lambda_b(v,s)\rangle = \,
\eta(w) \, \bar v_c (v',s') \Gamma v_b (v,s) .
\end{equation}
Hence we get
\begin{equation}
\eta(w) = \langle \phi (v')|\phi (v) \rangle.
\end{equation}

On the other hand, we can also evaluate the $Q_b \rightarrow Q_c$ matrix
element.
\begin{equation}
\langle Q_c (v_c',s_c') |\bar Q_c \Gamma Q_b|Q_b (v_b,s_b) \rangle =
\bar u_c (v_c',s_c') \Gamma u_b (v_b,s_b).
\end{equation}
However, since both ${\bf v_b}-{\bf v}={\bf p}/m_b$ and ${\bf v_c'}-{\bf v'}
={\bf p'}/m_c$ are small quantities of the order of $\Lambda_{\rm QCD}/m_Q$,
we can write
\begin{equation}
\langle Q_c (v_c',s_c') |\bar Q_c \Gamma Q_b|Q_b (v_b,s_b) \rangle \sim
\bar u_c (v',s_c') \Gamma u_b (v,s_b) + O(\Lambda_{\rm QCD}/m_Q).
\end{equation}
The subleading terms will be neglected in our discussion.

Lastly, making use of the fact that
\begin{equation}
\langle Q_a (v_a')|Q_a (v_a) \rangle = \delta^3(m_a{\bf v_a'}-m_a{\bf v_a})
= \delta^3({\bf p'}-{\bf p}-{\bf q} ),
\end{equation}
with ${\bf q}=m_a({\bf v'}-{\bf v})$, one of the integrals can be evaluated.
Then we get
\begin{eqnarray}
\langle T_{ac}(v',S',m')|\bar Q_c \Gamma Q_b|T_{ab} (v,S,m) \rangle &=&
\int d^3{\bf p} \, \psi (C;{\bf p+q}) \, \psi (B;{\bf p}) \, \delta_{s_as_a'}
\, \bar u_c (v',s_c') \Gamma u_b (v,s_b) \nonumber\\ & & \qquad \eta(w) \,
(\textstyle {1\over2},s_a; \textstyle {1\over2},s_b|S,m) \, (\textstyle
{1\over2},s_a'; \textstyle {1\over2},s_c'|S',m').
\end{eqnarray}

To see the significance of the integral, consider the semileptonic decay $Q_a
Q_bq \rightarrow Q_aQ_cq$ which is described by a single Isgur--Wise form
factor $\eta_{abc}(w)$ \cite{T4}.
The ``brown muck'' of a $QQq$ baryon is similar to that of a heavy meson $\bar
Qq$, with the two closely bound heavy quarks playing the role of the antiquark.
As the result, the analysis goes exactly as above except that the overlap of
the light degrees of freedom gives the mesonic Isgur--Wise form factor
${(w+1)\over2}\xi(w)$ instead of $\eta(w)$.
Setting the slow varying ${(w+1)\over2}\xi(w)$ to unity, we have
\begin{equation}
\langle Q_aQ_cq (v')|\bar Q_c \Gamma Q_b|Q_aQ_bq (v) \rangle = \int d^3{\bf p}
\, \psi (C;{\bf p+q}) \, \psi (B;{\bf p}) \, \bar u_c (v') \Gamma u_b (v).
\label{me}
\end{equation}
Hence, according to the definition in Ref. \cite{T4}, we conclude that
\begin{equation}
\eta_{abc}(w)=\eta_{abc}\left(1-{{\bf q}^2 \over 2m_a^2} \right) = \int d^3{\bf
p} \, \psi (C;{\bf p+q}) \, \psi (B;{\bf p}).
\end{equation}

Hence we finally obtain
\begin{eqnarray}
\langle T_{ac}(v'&,S',&m')|\bar Q_c \Gamma Q_b|T_{ab} (v,S,m) \rangle
\nonumber\\ &=& \eta(w) \, \eta_{abc}(w) \, \delta_{s_as_a'} \, \bar u_c
(v',s_c') \Gamma u_b (v,s_b) \, (\textstyle {1\over2},s_a; \textstyle
{1\over2},s_b|S,m) \, (\textstyle {1\over2},s_a'; \textstyle
{1\over2},s_c'|S',m').
\end{eqnarray}
Comparing with Eq. (\ref{def}), we get the main result of this paper.
\begin{equation}
\tilde \eta_{abc}(w) = \eta(w) \, \eta_{abc}(w).
\label{main}
\end{equation}
$\eta_{abc}(w)$ describes the perturbative attraction between the two heavy
quarks, while $\eta(w)$ accounts for the non-perturbative interaction between
the heavy quarks and the ``brown muck.''

It is in order to discuss the $w$-dependences of various form factors involved
in our discussion.
The meson Isgur--Wise form factor $\xi(w)$ is a slow varying function as its
slope at the point of zero recoil is of the order of unity.
On the other hand, $\psi(B,{\bf p})$ has a short range of the order of $B\sim
m_Q\alpha_s(m_Q)$.
Hence, near the point of zero recoil, the slope $\eta_{abc}(w)$ is of the order
of $\alpha_s^{-2}(m_Q)$.
In the heavy quark limit, this slope is large ($\alpha_s^{-2}(m_b) \sim 22.5$)
and the $w$-dependence of $\eta_{abc}(w)$ does overwhelm that of $\xi(w)$.
This justifies the neglect of the $\xi(w)$ factor in Eq. (\ref{me}).
On the other hand, we expect the baryon Isgur--Wise form factor $\eta(w)$ to be
slow varying in the real world.
In some models, however, $\eta(w)$ has strong $w$-dependence.
For example, in the large $N_c$ limit \cite{B1,B2}, the slope at the point of
zero recoil is of the order of $N_c^{3/2}$.
Hence we keep $\eta(w)$ in our final expression Eq. (\ref{main}) for
generality.

By Luke's Theorem \cite{B3}, $\eta(w)$ is normalized to unity in the heavy
quark limit.
\begin{equation}
\eta(1)=1.
\end{equation}
No analogous statement exists for $\eta_{abc}(w)$, the analytical form of which
is given in Ref. \cite{T4}.
Indeed it can be seen that
\begin{equation}
\eta_{abc}(1)=\left({2\sqrt{BC}\over B+C}\right)^3 ,
\end{equation}
which is not equal to unity unless $B=C$, i.e, $m_b=m_c$.
This is very different from the normalization of $\eta(w)$, which holds
regardless of the size of $m_b-m_c$ as long as both $m_b$ and
$m_c\gg\Lambda_{\rm QCD}$.

In conclusion we found that the semileptonic decays of heavy tetraquarks are
described by a single Isgur--Wise form factor $\tilde\eta_{abc}(w)$, which can
be factorized into two pieces.
The piece due to non-perturbative QCD is just the Isgur--Wise form factor
$\eta(w)$ for $Qqq \rightarrow Qqq$ decays, while the perturbative piece is
$\eta_{abc}(w)$, the Isgur--Wise form factor for $QQq  \rightarrow QQq$ decays.

The discussion above is expected to hold when $\alpha_s^2(m_Q)m_Q\gg
\Lambda_{\rm QCD}$.
In the real world, since the top quark does not live long enough to form
hadrons, we just got two ``hadronizable'' heavy quarks, the $b$-quark and the
$c$-quark.
The assumption above, however, holds for neither of them, and our results
cannot be applied directly.
Still it is possible that the picture above is at least qualitatively correct
and can serve as the starting point of quantitative investigations of the heavy
tetraquark systems by including the effects of $1/m_Q$ corrections.

\bigskip
This work was supported in part by the U.S. Dept. of Energy under Grant No.
DE-FG03-92-ER 40701.

\end{document}